\documentclass[fleqn,10pt]{SelfArx} 

\usepackage[english]{babel} 

\usepackage{natbib}
\usepackage{xspace}

\DeclareRobustCommand{\ion}[2]{%
\relax\ifmmode
\ifx\testbx\f@series
{\mathbf{#1\,\mathsc{#2}}}\else
{\mathrm{#1\,\mathsc{#2}}}\fi
\else\textup{#1\,{\mdseries\textsc{#2}}}%
\fi}

\definecolor{color1}{RGB}{0,0,90} 
\definecolor{color2}{RGB}{0,20,20} 

\usepackage{hyperref} 
\hypersetup{hidelinks,colorlinks,breaklinks=true,urlcolor=color2,citecolor=color1,linkcolor=color1,bookmarksopen=false,pdftitle={Title},pdfauthor={Author}}

\JournalInfo{White paper for Chandra cool attitude targets (CAT)} 

\PaperTitle{Stellar activity with TESS and Chandra} 
\Archive{}

\Authors{Hans Moritz G\"unther\textsuperscript{1}*, David A. Principe\textsuperscript{1}, Carl Melis\textsuperscript{2}, Kristina Monsch\textsuperscript{3}, P. Christian Schneider\textsuperscript{4}, S. Czesla\textsuperscript{4}, Nicholas J. Wright\textsuperscript{5}, Vinay L. Kashyap\textsuperscript{6}, J.H.M.M. Schmitt\textsuperscript{4},
E.R. Newton\textsuperscript{1}, Jeremy J. Drake\textsuperscript{6}, D. P. Huenemoerder\textsuperscript{1}}
\affiliation{\textsuperscript{1}\textit{Massachusetts Institute of Technology}} 
\affiliation{\textsuperscript{2}\textit{UC San Diego}} 
\affiliation{\textsuperscript{3}\textit{Universit\"ats-Sternwarte, Ludwig-Maximilians-Universit\"at M\"unchen}}
\affiliation{\textsuperscript{4}\textit{Hambuger Sternwarte}}
\affiliation{\textsuperscript{5}\textit{Keele University}}
\affiliation{\textsuperscript{6}\textit{Harvard-Smithsonian Center for Astrophysics}}
\affiliation{*\textbf{Corresponding author}: hgunther@mit.edu} 

\Keywords{} 
\newcommand{\keywordname}{Keywords} 

\Abstract{All cool stars show magnetic activity, and X-ray emission is the hallmark of this activity. Gaining an understanding of activity aids us in answering fundamental questions about stellar astrophysics and in determining the impact of activity on the exoplanets that orbit these stars. Stellar activity is driven by magnetic fields, which are ultimately powered by convection and stellar rotation. However, the resulting dynamo properties heavily depend on the stellar interior structure and are far from being understood.
X-ray radiation can evaporate exoplanet atmospheres and damage organic materials on the planetary surface, reducing the probability that life can form or be sustained, but also provides an important source of energy for prebiotic chemical reactions. 
Over the next two years, the TESS mission will deliver a catalog of the closest exoplanets, along with rotation periods and activity diagnostics for millions of stars, whether or not they have a planet.
We propose to include all cool stars that are TESS targets and bright enough for Chandra observations, as determined by their detection and flux in the ROSAT all-sky survey (RASS), to the list of Chandra Cool Attitude Targets (CATs). 
For each target, the signal will be sufficient to fit the coronal plasma with at least two temperature components, and compare abundances of groups of elements with low, medium, or high first ionization potential. Similar to the known relation between X-ray luminosity $L_X$ and rotation period, we can correlate stellar properties with coronal temperatures and abundances to constrain models for stellar activity, coronal heating, and stellar dynamos.  Detailed X-ray characterization for even a subset of planet-hosting systems would dramatically advance our knowledge of what impact these emissions have on orbiting planets.
}

\begin{document}

\flushbottom 

\maketitle 


\thispagestyle{empty} 


\section{Introduction}
Most stars in the universe show X-ray activity. For cool, low-mass stars ($<$3 M$_\odot$) this is powered by a magnetic dynamo and for high-mass stars, shocks in the wind produce X-rays; there is a small gap between these two mechanisms which leaves early A-type stars X-ray dark \citep[see review by][]{2009A&ARv..17..309G}. X-ray observations can probe stellar activity, the underlying magnetic dynamo, as well as determine stellar X-ray properties (flux, temperature, flare rate) that are crucial for understanding the evolution of circumstellar material, from young accretion disks to exoplanets, and the development of life on those planets. 

\subsection{Stellar activity}

\begin{figure*}[ht]\centering
\includegraphics[width=\linewidth]{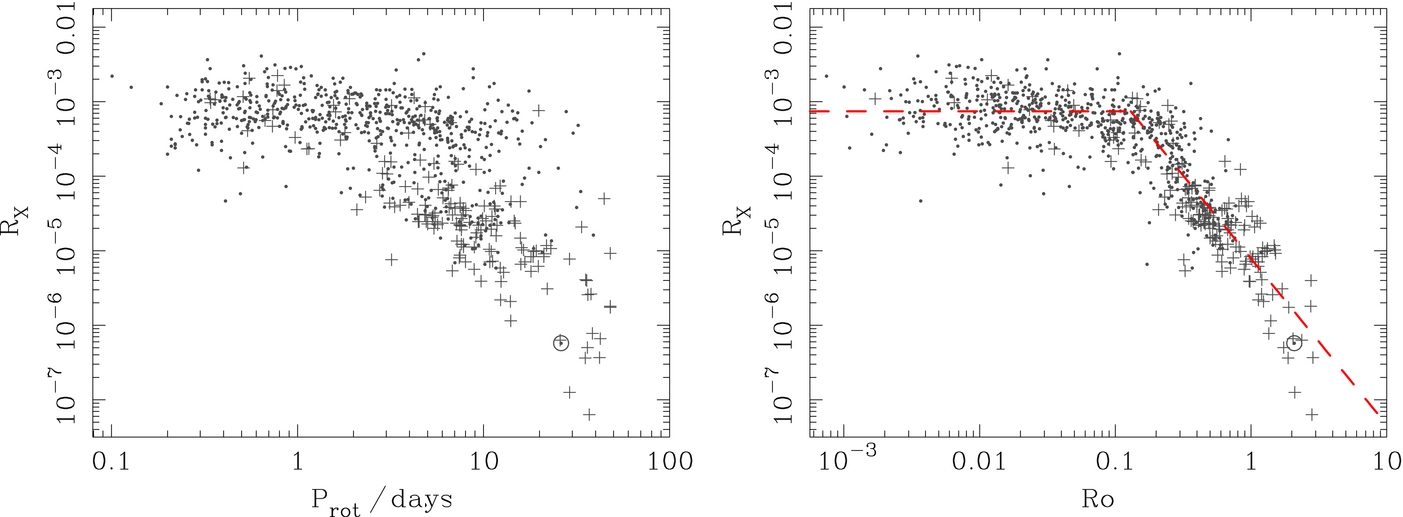}
\caption{$L_X/L_{\mathrm{bol}}$ depending on rotation period (left) or Rossby number (right). Known binaries are marked as ``+'' and the Sun is indicated at the bottom right with a circle. Figure from \protect{\citet{Wright2011}}.}
\label{fig:wright}
\end{figure*}

Stellar activity is best studied in our own Sun, where we observe an activity cycle lasting about 11~years. 
This cycle was first discovered by simply counting the number of daily Sun-spots.
Other tracers of activity seen in both the Sun and other stars include the X-ray flux, the number of flares, or the so-called S-index that characterizes emission in the Ca~H and K lines. This `stellar activity' is closely connected to the star's magnetic field and driven by a dynamo which 
is powered by differential rotation in the stellar interior. We are still lacking a dynamo theory that can quantitatively predict details like the length of a stellar cycle. Observational data beyond our Sun is relatively sparse and contributes to our lack of understanding of the origins of stellar activity. We know stellar activity is closely correlated with the rotation period $P_{\mathrm{rot}}$, or more precisely, the Rossby number, $Ro = P_{\mathrm{rot}}/\tau$, where $\tau$ is the convective turn-over time of the stellar atmosphere ($\tau$ depends on the mass of the star)---see Fig.~\ref{fig:wright}. For slowly rotating stars the ratio of X-ray luminosity $L_X$ and bolometric luminosity $L_{\mathrm{bol}}$ increases with decreasing $P_{\mathrm{rot}}$. \citet{Wright2011} compiled a sample of about 800 stars with known X-ray fluxes and rotation periods and found $L_X/L_{\mathrm{bol}} \sim Ro^\beta$ with $\beta = -2.70 \pm 0.13$. They interpret this number as a sign that stars rotating more slowly have less differential rotation. 
At $\log (L_X/L_{\mathrm{bol}})\sim-3$ the X-ray flux saturates, possibly due to a different configuration of the dynamo \citep{Wright2011}. There is evidence that the fractional X-ray luminosity may decrease for the most rapid rotators (super-saturation), though observational evidence for this is sparse.

\subsection{Stellar activity with TESS}

The Transiting Exoplanet Survey Satellite \citep[TESS,][]{TESS} is aimed at detecting transiting exoplanets in the solar neighborhood. However, this detection method will also produce a wealth of data on stellar activity. TESS is obtaining high-precision photometry of $>200,000$ stars with a cadence of 2 minutes; all remaining sources in the field-of-view will have lightcurves with 30~min cadence. Unlike CoRoT, \emph{Kepler}, and K2, TESS will observe most of the sky. ``Sectors'' of 27.4 day monitoring periods are being conducted, first in the southern ecliptic hemisphere over the course of one year, followed by the northern hemisphere in the second year. 

\textbf{TESS will give us rotation periods for stars with rapid periods, constraints on the period for stars with slow periods, and flare statistics on each target.}
Given TESS's small aperture relatively bright sources are required, but most targets in this program are within 100~pc and fullfil this criterion (see Sect.\ref{sect:feasibility}). 
For active stars, we can expect many flares within the TESS monitoring period. For example, GJ~1245 A \& B flare several times per day and \citet{Lurie15} used \emph{Kepler} data of this system to constrain stellar age-rotation-activity models. Other examples of \emph{Kepler} or K2 data important for understanding stellar activity are sample studies of superflares \citep[e.g.][]{Shibayama13}, rotation periods \citep[e.g.][]{Douglas16}, or flares on M dwarfs \citep{Hawley14}, as well as numerous studies of individual systems \citep[e.g.][]{Davenport14}. 
The fast cadence lightcurves obtained by TESS will allow similar studies, {\bf but unlike \emph{Kepler}, most TESS fast cadence targets are close to the Sun and thus amenable to coronal studies from fairly short Chandra observations}.

\section{Science questions}
We propose to add all close-by, bright stars as CAT targets (details in Sect.~\ref{sect:targets}). Any random subset of these targets will provide significant opportunities to study stellar activity in more detail than ever before (Sect.~\ref{sect:activity}); however, this data will also provide a legacy enabling other studies (possibly in conjunction with data that does not yet exist such as the TESS list of detected planets or the \emph{GAIA} data on binary orbits). In Sect.~\ref{sect:planets} and \ref{sect:binaries} we give examples for more detailed questions that could be answered with the proposed CAT targets.

\subsection{Stellar activity}
\label{sect:activity}

Most work on samples of stellar activity has been done with \emph{ROSAT} \citep[e.g.][]{Pizzolato2003,2004A&A...417..651S}.
\citet{Wright2011} collected $>800$ stars with X-ray data and rotation periods (Fig.~\ref{fig:wright}), including some \emph{XMM-Newton} observations, but primarily based on \emph{ROSAT} data. The X-ray information was therefore limited to the flux, with no additional information such as estimates of plasma temperature or abundances. However, a spectrum can reveal a much more detailed picture of the corona than a simple count rate. For example, \citet{Preibisch2005} fit a two-temperature model to the young stars in the Orion Nebula Cluster and study how the two temperatures are related to each other and to other stellar properties.From studies of individual low-activity stars, we can identify a FIP (first ionization potential) effect where elements of low FIP such as Fe are enhanced and elements of high FIP like Ne are depleted, while the opposite is true in stars of high activity level (IFIP) - see \citet{2009A&ARv..17..309G}. The coronal Ne abundance is an open question even in our Sun which leads to significant tension between measured solar abundances in spectroscopy and the requirements of helioseismology \citep[see][for a detailed discussion]{Drake2011}.

An additional problem is that the $L_X$ of stars can vary significantly over a stellar cycle; for our Sun the typical X-ray flux varies by a factor of 10 between solar maximum and minimum and X-ray cycles have also been observed on some other stars \citep[e.g.][]{Robrade12}. 
Comparison of the X-ray flux between the original \emph{ROSAT} observation and new Chandra observations would allow us to constrain the level of X-ray variability (whether due to flaring or activity cycles) on $\sim$25 year timescales 
\citep[see, e.g.,][]{1999ApJ...524..988K}.
This would allow us to determine how much of the spread in the rotation-activity relationship (Fig.~\ref{fig:wright}) is due to variability and how much is inherent in the relationship.

\citet{2017xru..conf..326S} cross-matched all \emph{Kepler} lightcurves with the \emph{XMM-Newton} source catalog. They find 107 X-ray emitters, but given the average distance of their \emph{Kepler} targets of 270~pc, they are limited to the most active stars and the X-ray data is limited to just count rates, not spectra, let alone detailed plasma temperature or coronal abundances. In the sample we propose here, all targets are located much closer to the Sun and thus their X-ray flux is much higher, and other stellar properties are also easier to obtain from ground-based studies.

A number of close-by main-sequence stars have been studied in great detail with \emph{XMM-Newton} and \emph{Chandra}, often using the gratings, but the total sample size remains small ($<50$ stars in \citealt{Ness2004} and $<20$ in \citealt{Wood2018}). These studies show the potential science return of detailed spectroscopy of stellar coronae, and the CAT program is an ideal opportunity to multiply the sample size and enable these and other studies by the community in the future. Particular examples are:
\begin{itemize}
    \item Fit two or more temperature components and study how the temperature depends on $P_{\mathrm{rot}}$ or stellar mass to constrain dynamo models,
    \item fit abundances of low, medium, and high FIP elements to constrain models of elements differentiation in the corona, 
    \item compare optical flare rates to the quiescent X-ray flux.
\end{itemize}

A \emph{Chandra} pointing of 10~ks is long enough to check whether the star is observed in quiescence or during a flare, so that observations of flaring stars can be treated specially to reduce the scatter in these relations.

\subsection{The impact of stellar activity on planets}
\label{sect:planets}

Stellar activity affects the formation, atmospheric evolution, and habitability of any planetary companions.
Atmospheres can be photoevaporated, surfaces can be sublimated, life
can be stymied or in some cases promoted by enhanced activity levels
\citep[e.g.,][]{Owen+2010, Lammer2003, AlexanderPascucci2012, QuintanaLissauer2014, OwenWu2017, Mullan2008, Ranjan2017}.
For mature planetary systems, precisely reconstructed X-ray and extreme ultraviolet radiative input from the host star is essential in determining 
the potential impact on their planets (e.g., \citealt{Rugheimer2015, France2016, louden17}). Unfortunately, most of the ultraviolet emission is not directly observable due to absorption by interstellar material. 

Combining X-ray and ultraviolet (e.g., GALEX, IUE, FUSE, and/or HST) data can provide reasonable constraints on expected extreme ultraviolet emission levels, thus informing models exploring the evolutionary history and fate of exoplanets on a case-by-case basis. Such analysis requires detailed X-ray spectroscopy that is robust enough to have its models extrapolated over the extreme ultraviolet wavelength range to meet with existing ultraviolet data.
\textbf{Increasing the sample of exoplanet-hosts that are observed in X-rays can therefore strongly help in constraining the formation, orbital and atmospheric evolution, and habitability of planets around nearby stars.} 

\subsection{Binaries}
\label{sect:binaries}
With \emph{Chandra}'s high angular resolution capabilities known X-ray emitting visual binaries with angular separations between 1" - 20" can be resolved, which are blended in \emph{ROSAT}, \emph{XMM-Newton}, or e-ROSITA. 
Physical binaries should share the same age, abundances, and evolutionary histories and thus provide a good laboratory to study how X-ray activity depends on these quantities.

\section{Target selection criteria}
\label{sect:targets}
\begin{figure}[ht]\centering
\includegraphics[width=0.8\linewidth]{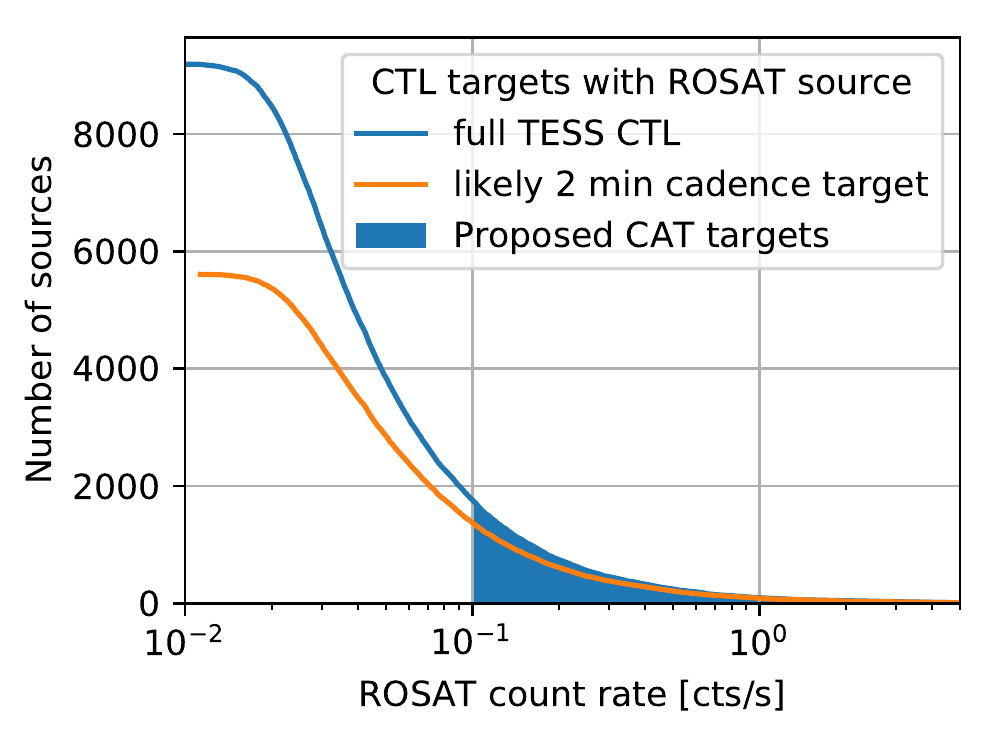}
\caption{Number of sources in the CTL with a ROSAT counterpart at least as bright as the value given on the x-axis.}
\label{fig:sourcenumber}
\end{figure}

 The TESS Candidate Target List \citep[CTL,][]{2018AJ....156..102S} essentially contains all sources with a known parallax or proper motion (i.e.\ galactic point sources) that are bright enough for TESS. The CTL contains mostly dwarfs within a few hundred pc of the Sun. We remove all targets outside the ecliptic latitude range allowed for CATs. The CTL also prioritizes targets based on the probability to detect a transiting planet. The top few $10^5$ targets on this list will be observed with a 2 min cadence. We cross-match the CTL with the Second ROSAT all-sky survey (2RXS) source catalog \citep{2016A&A...588A.103B} allowing for a maximum separation of 0.5~arcmin (a typical 90\% uncertainty for ROSAT observations). Figure~\ref{fig:sourcenumber} shows the number of sources in the merged catalog that have a count rate higher than a certain brightness. Most of the targets bright in the 2RXS also have a high priority to be observed with a 2~min cadence in TESS.

\section{Feasibility: Exposure time and instrument choice}
\label{sect:feasibility}
\begin{figure}[ht]\centering
\includegraphics[width=0.8\linewidth]{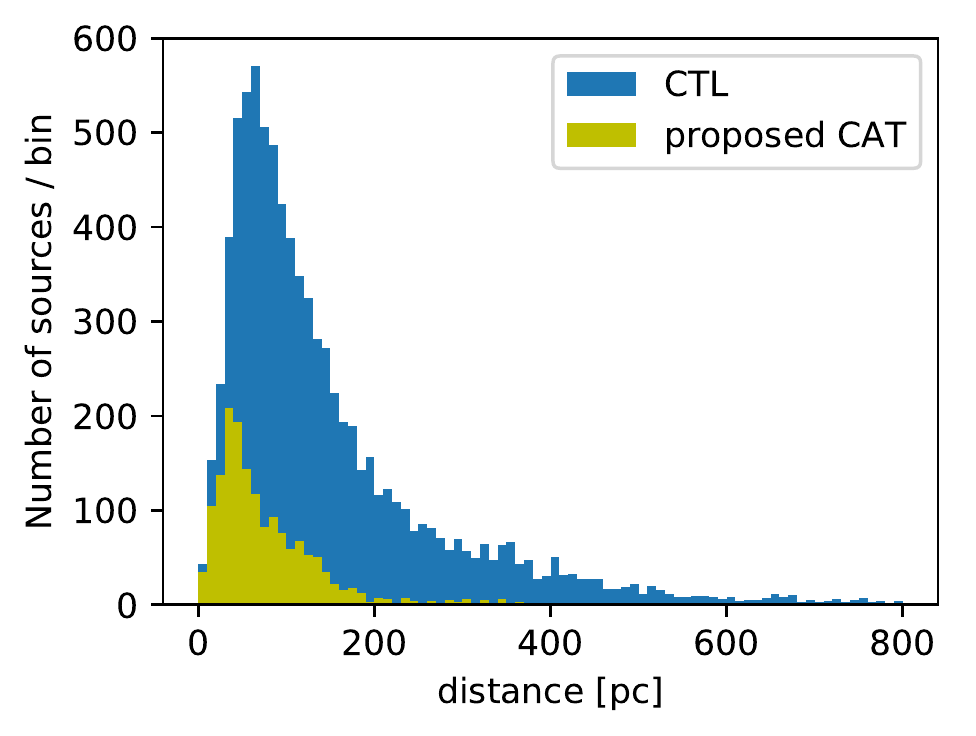}
\caption{Distance to sources in our input list.}
\label{fig:sourcedistance}
\end{figure}

Our science goal requires us to obtain a spectrum, ruling out imaging observations with the HRC. Since stars are typically fairly soft targets, imaging should be done with the ACIS-S3 chip. For the brightest targets we suggest spectroscopy with the LETG/HRC-S.
Most of our targets are located within 100~pc (Fig.~\ref{fig:sourcedistance}), so that the galactic column density is low. The LETG/HRC-S includes tracers of cool coronal components that would be undetectable with the HETGS.

The relation between the 2RXS count rate and the Chandra count rate depends on the properties of the emitting plasma. Hotter plasma produces harder spectra and higher count rates in Chandra/ACIS. The 2RXS contains fitted parametric models, including optically thin emission from coronal sources, but unfortunately most values are so uncertain that they are of little use to predict the Chandra count rates; even the hardness ratios are very uncertain for sources that were only observed for a few hundred seconds in the all-sky survey. For coronal temperatures of 0.2, 0.5, 1.0, and 2.0~keV, the ACIS-S count rate would be 0.1, 0.4, 0.6, and 0.9 times the 2RXS count rate according to WebPIMMS assuming an APEC plasma model with solar abundances and no galactic $N_H$. The science described above requires (a) a lightcurve with $>50$ counts per bin and at least ten bins to distinguish quiet targets (flat lightcurve) from targets that flare during the observation (rising or falling lightcurves) and (b) the fit of a two-temperature, variable abundance plasma model. \citet{Guenther2017} analyze a source with about 1000 counts on ACIS-S3. The temperature of each of the two components carries about a 20\% uncertainty. Fixing the oxygen abundance to 1, the abundance fitted for a group of low-FIP elements (Mg, Fe, Si) is $0.2\pm0.1$, while Ne (high FIP) gives $1.2^{+0.4}_{-0.6}$. This is sufficient to determine if a FIP or IFIP effect is present. Following this example, we require a minimum of 1000 counts. 
Assuming an ACIS-S count rate that is 0.5 times the 2RXS rate, we suggest the following strategy: \textbf{Observe all targets with a 2RXS rate $>0.2$~cts/s for 10~ks and those with an 2RXS rate between 0.1 and 0.2~cts/s for 20~ks.} Figure~\ref{fig:sourcenumber} shows that there are about 2000 targets with a 2RXS rate $>0.1$~cts/s.

The LETG/HRC-S effective area peaks in the same energy range as ROSAT/PSPC and the conversion rate between the 2RXS count rate and the predicted LETG/HRC-S first order count rate is about 0.1 for all temperatures. We suggest observing the brightest $\sim$200 sources (2RXS count rate $>0.25$) with the LETG/HRC-S. A 35~ks exposure will provide about 1000 counts in the spectrum, enough to identify about a dozen emission lines for a detailed temperature and abundance diagnostic \citep[compare data in][]{Wood2018}. 

\section{Target priority}
We suggest about 2000 CAT targets. Of those, targets where Chandra can provide information that is not available from ROSAT, the XMM slew survey, or the upcoming e-ROSITA all sky survey shall be prioritized. This includes the 200 brightest sources for LETG/HRC-I spectroscopy and a few hundred known visual binaries that \emph{Chandra} can resolve. Since we are concerned with stellar activity and variability, repeat observations of targets already in the \emph{Chandra} archive are particularly valuable, so targets with existing data shall also be prioritized.




\phantomsection
\bibliographystyle{prop}
\setlength\bibsep{0pt}
\bibliography{references}


\end{document}